\documentclass[amsmath,amssymb,superscriptaddress,showpacs,preprint,floatfix]{revtex4}
\usepackage{graphicx}
\usepackage{latexsym}
\usepackage{amsmath}
\usepackage{amsfonts}
\usepackage{amssymb}

\begin{document}

%

\title{Generalized Schr\"odinger cat states and their classical emulation}

\author{Armando Perez-Leija}
\address{Institute of Applied Physics, Abbe Center of Photonics, Friedrich-Schiller-Universit\"at Jena, Max-Wien-Platz 1, 07743 Jena, Gemany}\author{Ir\'an Ramos-Prieto}\address{Instituto Nacional de Astrof\'{i}sica, \'{O}ptica y Electr\'{o}nica, Calle Luis Enrique Erro No. 1, Santa Maria Tonantzintla, Pue. CP 72840, M\'{e}xico}
\author{Alexander Szameit}\address{Institute of Applied Physics, Abbe Center of Photonics, Friedrich-Schiller-Universit\"at Jena, Max-Wien-Platz 1, 07743 Jena, Gemany}
\author{Demetrios N. Christodoulides}\address{CREOL/College of Optics, University of Central Florida, Orlando, Florida 32816, USA}
\author{Hector Moya-Cessa}\address{Instituto Nacional de Astrof\'{i}sica, \'{O}ptica y Electr\'{o}nica, Calle Luis Enrique Erro No. 1, Santa Maria Tonantzintla, Pue. CP 72840, M\'{e}xico}

\begin{abstract}
We demonstrate that superpositions of coherent and displaced Fock states, also referred to as generalized Schr\"odinger cats cats, can be created by application of a nonlinear displacement operator which is a deformed version of the Glauber displacement operator. Consequently, such generalized cat states can be formally considered as nonlinear coherent states. We then show that Glauber-Fock photonic lattices endowed with alternating positive and negative coupling coefficients give rise to classical analogs of such cat states. In addition, it is pointed out that the analytic propagator of these deformed Glauber-Fock arrays explicitly contains the Wigner operator opening the possibility to observe Wigner functions of the quantum harmonic oscillator in the classical domain.
\end{abstract}


\maketitle



\section{Introduction}

Since the invention of the laser, the generation and manipulation of quantum sates of light have been subjects of extensive investigations \cite{Haroche,Wineland1,Silberberg}. In quantum optics, noise fluctuations of coherent states (CS) are referred to as the quantum limit and therefore they establish the boundary between classical and quantum sates \cite{Gerry,Heilmann}. Along those lines, there exist several nonclassical states which exhibit lower noise fluctuations than CS. Among them one may mention squeezed states \cite{Gerry} and nonlinear CS \cite{Manko}. These latter states are a generalization of the standard CS originally introduced by Glauber \cite{Glauber1, Glauber2}. In this regard, Man'ko {\it et al.} \cite{Manko} introduced nonlinear CS as the eigenstates of a deformed annihilation operator $\hat{A}=\hat{a}f(\hat{n})$, where $f\left(\hat{n}\right)=f\left(\hat{a}^{\dagger}\hat{a}\right)$ represents an arbitrary function of the number operator of the harmonic oscillator, and $\hat{a}^{\dagger}$ and $\hat{a}$ are the creation and annihilation operators, respectively. Accordingly, nonlinear CS are generated from the vacuum state by the action of the nonlinear displacement operator $\hat{D}_{NL}\left(\alpha\right)=\exp(\alpha \hat{A}^{\dagger}-\alpha^*\hat{A})$ 

\begin{equation}\label{eq:1}
|\alpha\rangle_{NL}=\hat{D}_{NL}\left(\alpha\right)|{0}\rangle.
\end{equation}
where $\alpha$ is a complex number. More generally, when considering Fock states, $|{k}\rangle$, as initial states in Eq. (\ref{eq:1}), we expect to produce nonlinear displaced Fock states (DFS). The aim of this work is to show that a particular choice of the function $f(\hat{n})$ for the operators $\hat{A}$ and $\hat{A}^{\dagger}$ give rise to superpositions of either CS or DFS, depending on whether the vacuum or a number state is initially excited. In the quantum optics literature, superpositions of CS are referred to as Schr\"odinger cat states \citep{Gerry}. At the same time DFS are considered a generalization of CS, and in that vein superpositions of DFS can be considered a generalization of Schr\"odinger cat states \cite{Gerry,Schrodinger,Yurke}.

As a second part of our work, we describe the possibility to optically emulate such cat states using a deformed version of the so-called Glauber-Fock photonic lattices \cite{GF}. 
Furthermore, we show that the analytic propagator of the deformed Glauber-Fock lattices can be written in terms of the Glauber displacement operator and the Wigner operator \citep{louisell}. As a result, light fields propagating through these types of lattices will feature properties akin to Wigner functions of the quantum harmonic oscillator. 

\section{Generalized Schr\"{o}dinger cat states as Nonlinear coherent states}
\label{sec:NCS}

In order to generate nonlinear coherent states directly from Eq. (\ref{eq:1}), we consider  the deformed annihilation and creation operators

\begin{equation}\label{eq:2}
\hat{A}= (-1)^{\hat{n}}\hat{a}, \qquad \hat{A}^{\dagger} =\hat{a}^{\dagger}(-1)^{\hat{n}},
\end{equation}
with $(-1)^{\hat{n}}$ representing the photon-number parity operator \cite{Gerry,Gerry2}. After a close inspection of these operators one finds that the commutator $[\hat{A},\hat{A}^{\dagger}]=1$. This in turn allows us to disentangle the nonlinear displacement operator $\hat{D}_{NL}\left(\alpha\right)$ according to the Baker-Hausdorff formula \cite{louisell}
\begin{equation}\label{eq:3}
\hat{D}_{NL}\left(\alpha\right)=\exp\left(-|\alpha|^2/2\right)\exp\left(\alpha \hat{A}^{\dagger}\right) \exp\left(-\alpha^*\hat{A}\right).
\end{equation}

Expanding in Taylor the exponentials
\begin{eqnarray}\label{eq:4}
\exp\left(-\alpha^{*}\left(-1\right)^{\hat{n}}\hat{a}\right)=\cos\left(\alpha^{*} \hat{a}\right)+\sin\left(\alpha^{*} \hat{a}\right)\left(-1\right)^{\hat{n}},\nonumber\\
\exp\left(\alpha \hat{a}^{\dagger}\left(-1\right)^{\hat{n}}\right)=\cos\left(\alpha \hat{a}^{\dagger}\right)+\sin\left(\alpha \hat{a}^{\dagger}\right)\left(-1\right)^{\hat{n}},\end{eqnarray}
one can show that the displacement operator Eq. (\ref{eq:3}) can be written in terms of the usual Glauber displacement operator $\hat{D}\left(i\alpha\right)$ \cite{Gerry} 
\begin{eqnarray}\label{eq:5}
\hat{D}_{NL}\left(\alpha\right)=\frac{1}{2i}\left[\hat{D}\left(i\alpha\right)-\hat{D}^{\dagger}\left(i\alpha\right)\right]\left(-1\right)^{\hat{n}}\nonumber\\
+\frac{1}{2}\left[\hat{D}\left(i\alpha\right)+\hat{D}^{\dagger}\left(i\alpha\right)\right].
\end{eqnarray}
Operating with $\hat{D}_{NL}\left(\alpha\right)$ on a system prepared in a pure Fock state $|{k}\rangle$, we drive the system to evolve into a superposition of two DFS separated in phase
\begin{eqnarray}\label{eq:6}
|{\alpha}\rangle_{NL}=\frac{\exp\left(-i\left(-1\right)^{k}\pi/4\right)}{\sqrt{2}}|{i\alpha,k}\rangle\nonumber\\+\frac{\exp\left(i\left(-1\right)^{k}\pi/4\right)}{\sqrt{2}}|{-i\alpha,k}\rangle.
\end{eqnarray}
For clarity, we remind that DFSs are obtained by applying the Glauber displacement operator to any number state: $\hat{D}\left(\alpha\right)|{k}\rangle=|{\alpha,k}\rangle$ \cite{Gerry}. In this manner, for the particular case when the system starts in the vacuum state, $|{0}\rangle$, the system develops into a superposition of two CSs
\begin{eqnarray}\label{eq:6a}
|{\alpha}_{NL}\rangle=\frac{\exp\left(-i\pi/4\right)}{\sqrt{2}}|{i\alpha}\rangle\nonumber\\+\frac{\exp\left(i\pi/4\right)}{\sqrt{2}}|{-i\alpha}\rangle.
\end{eqnarray}
On the other hand, the action of $\hat{D}_{NL}\left(\alpha\right)$ over a DFS, $|{\psi(0)}\rangle=|{\beta,k}\rangle$, forces the system to evolve into a  superposition of four DFSs 
\begin{eqnarray}\label{eq:7}
&\hat{D}_{NL}\left(\alpha\right)|{\beta,k}\rangle=\exp\left(-iRe\left(\alpha\beta^{*}\right)\right)\left[|{-i\alpha+\beta,k}\rangle-i\left(-1\right)^{k}|{i\alpha-\beta,k}\rangle\right]/2\nonumber\\&+\exp\left(iRe\left(\alpha\beta^{*}\right)\right)\left[|{i\alpha+\beta,k}\rangle+i\left(-1\right)^{k}|{-i\alpha-\beta,k}\rangle\right]/2,
\end{eqnarray}
where $Re\left(x\right)$ indicates real part. Correspondingly, when the system starts in a CS, $|{\psi(0)}\rangle=|{\beta}\rangle$, it is straightforward to see that Eq. (\ref{eq:7}) reduces to a superposition of four CSs. These results demonstrate the possibility of generating high order superpositions of DFSs as well as CSs by constructing a deformed version of the Glauber displacement operator $\hat{D}_{NL}\left(\alpha\right)$.
Moreover, since superpositions of coherent as well as displaced Fock states are created by distorting the usual creation and annihilation  operators, these superposition of states, Eq. (\ref{eq:6a}) and Eq. (\ref{eq:7}), can be formally considered as nonlinear coherent states \cite{Manko}. For a full analysis and description of nonlinear CSs we refer to \cite{Manko}.  
\\ 
Before considering the classical emulation of Schr\"odinger cat states, it is important to note that deformations implemented via the parity photon-number operator, $(-1)^{\hat{n}}$, could be realized experimentally using schemes from quantum optical metrology \cite{Gerry2}. 
\begin{figure}
\centerline{\includegraphics[scale=0.3]{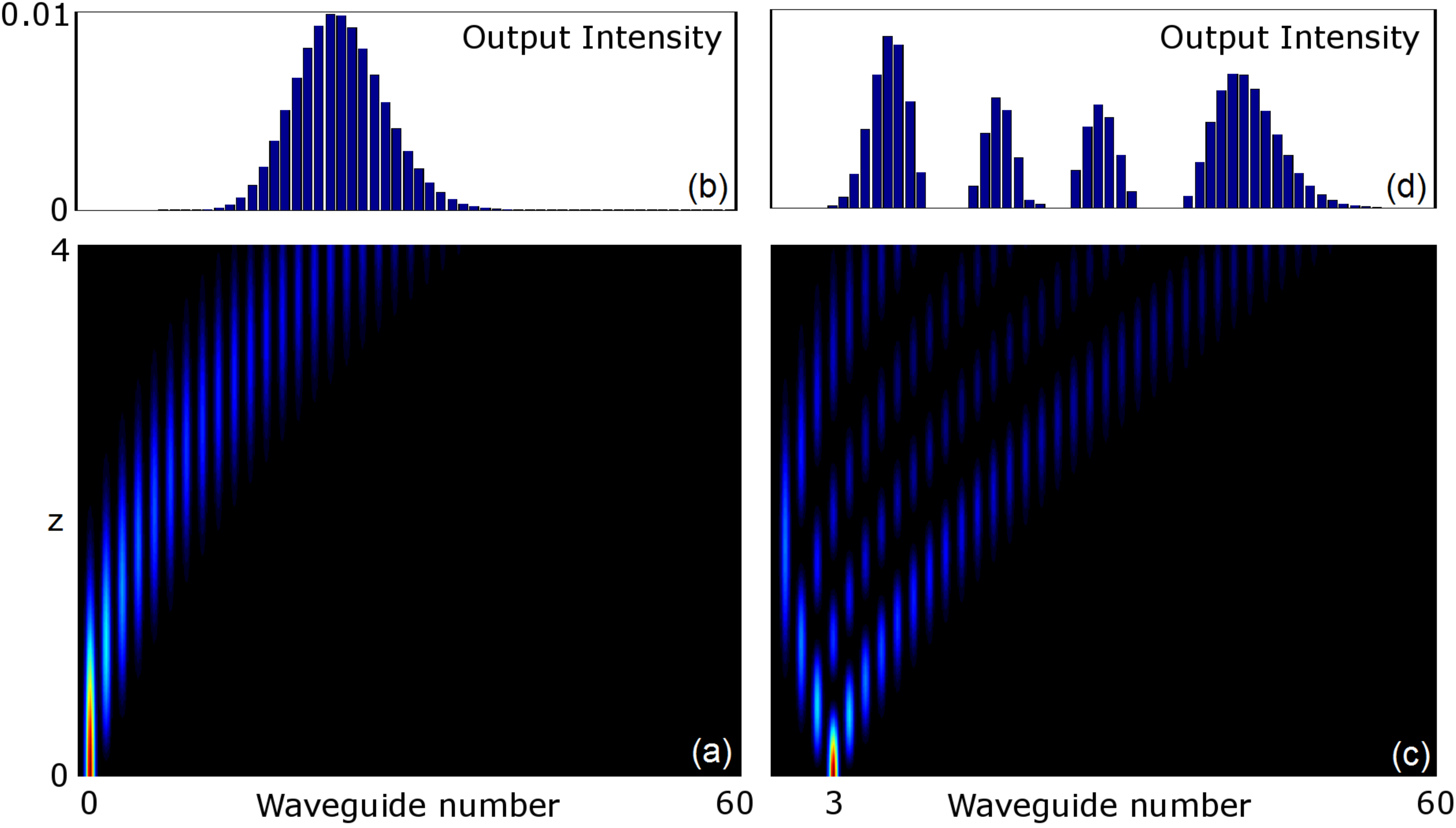}}
\caption {Theoretical results for single input excitation using a Glauber-Fock lattice of 60 waveguides and parameter $g=1$. Intensity evolution when the zeroth (a) and third (c) waveguides are excited. In these photonic arrays the states of the harmonic oscillator are mapped onto the lattice sites, and the lattice itself plays the role of the deformed displacement operator. As a result, the field distributions can be considered as the classical analogs of the Schr\"odinger cat states (left) and the third generalized Schr\"odinger cat state (right) of the quantum harmonic oscillator.}\label{Fig1}
\end{figure}
\section{Classical emulation of Schr\"odinger cat states using Glauber-Fock photonic lattices}
In this section we show how to model generalized Schr\"odinger cat states using waveguide arrays. 
In general, classical and non-classical light beams traversing periodic and nonperiodic waveguide lattices exhibit discrete diffraction which is a direct byproduct of the optical tunneling occurring between different lattice sites \cite{Christodoulides,Review,Garanovich}. Hence, by controlling the evanescent coupling between adjacent waveguide elements, as well as the local propagation constants, one can mold the light dynamics in order to perform specific physical processes \cite{Wstates}. 
In this regard, many innovative waveguide schemes have been implemented in order to classically emulate interesting quantum phenomena \cite{Longhi,GF2}. These are for instance displaced Fock states \cite{GF3}, PT-symmetric quantum systems \cite{PT}, observations of geometric phases \cite{Kai}, and massless Dirac particles \cite{Zeuner}. In the latter case, an optical analogue of relativistic massless Dirac particles was accomplished by using uniform waveguide lattices with alternating positive and negative coupling coefficients. To our knowledge, periodic waveguide structures with negative coupling coefficients were first considered theoretically in a classical context by Efremidis  \cite{Efremidis}.\\
Concurrently, in the quantum realm waveguide configurations have also played an important role in the development of modern quantum communication schemes \cite{Metcalf}, sensing systems \cite{Heilmann}, as well as for engineering photonic quantum circuits \cite{Wstates2} to achieve perfect transfer of quantum states \cite{Jx1,Jx2}, quantum fractional Fourier transforms \cite{Fourier}, and quantum random walks \cite{Peruzzo}.\\ 
\begin{figure}[t]
\centerline{\includegraphics[scale=0.3]{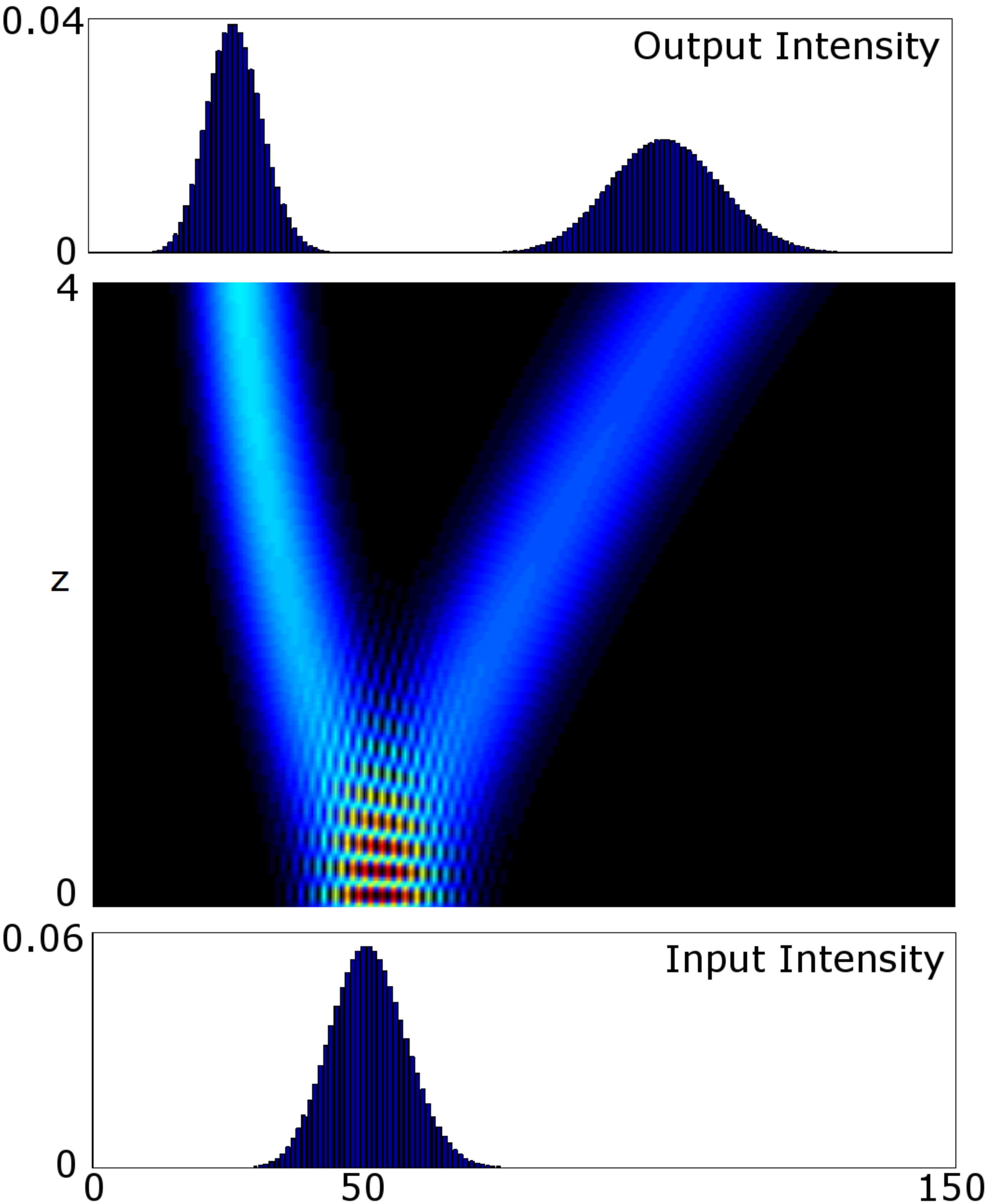}}
\caption {Evolution of a coherent state $|{\beta}\rangle$, with average photon number $\langle \hat{n}\rangle=50$, in a deformed Glauber-Fock lattice having 150 waveguides and $g=1$. A coherent state, which is described by a Poissonian intensity distribution, splits into two Poisson distributions along evolution. From Eq. (\ref{eq:14}) we infer that each distribution is composed by two out of phase Poisson distributions.}\label{fig:Fig2}
\end{figure}
In order to show how to emulate superpositions of CS and DFS using photonic  waveguide lattices, we consider the Schr\"odinger equation    
\begin{eqnarray}\label{eq:8}
i\frac{d|{\psi(z)}\rangle}{dz}+\hat{H}|{\psi(z)}\rangle=0,
\end{eqnarray}
with the Hamiltonian given by $\hat{H}=g(\hat{a}^{\dagger}(-1)^{\hat{n}}+(-1)^{\hat{n}}\hat{a})$, $g$ being an arbitrary constant, $(-1)^{\hat{n}}$ the photon-number parity operator \cite{Gerry}, and $z$ is the propagation distance.
Hence, expanding the wave function $|{\psi(z)}$ as a superposition of number states, $|{\psi(z)}\rangle=\sum_{n=0}^{\infty}E_n(z)|n\rangle$, using the properties of the annihilation and creation operators,
$\hat{a}|{n}\rangle=\sqrt{n}|{n-1}\rangle$ and $\hat{a}^{\dagger}|{n}\rangle=\sqrt{n+1}|{n+1}\rangle$, and the orthonormality of the number states, $\langle m|{n}\rangle=\delta_{m,n}$, we obtain a semi-infinite set of coupled differential equations for the transition probability amplitudes 
\begin{figure}[t]
\centerline{\includegraphics[scale=0.4]{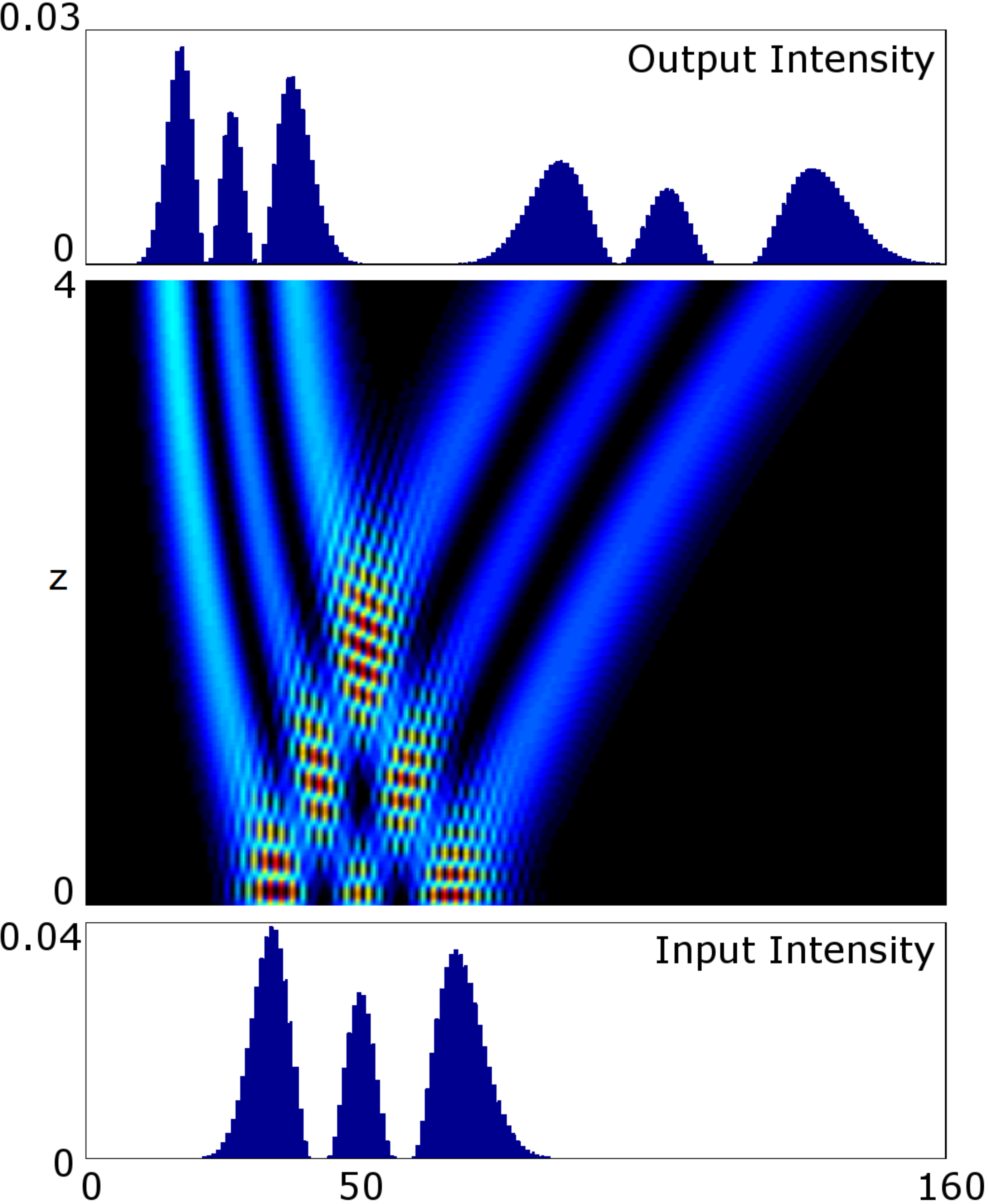}}
\caption{ Bottom: initial intensity profile corresponding to the second displaced Fock-state $|{\beta,2}\rangle$ with an "average photon-number" of $\langle \hat{n}\rangle=50$. Center: intensity evolution of such classical displaced Fock-state. Top: Output intensity showing a superposition of DFS. The evolution is calculated using a deformed Glauber-Fock lattice of 4cm long with 160 waveguides and $g=1$.}\label{fig:Fig3}
\end{figure}
\begin{eqnarray}\label{eq:9}
&& i\frac{dE_0}{dz}+gE_1=0, \nonumber \\
&& i\frac{dE_m}{dz}+g(-1)^m\left(\sqrt{m+1}E_{m+1}-\sqrt{n}E_{m-1}\right)=0,
\end{eqnarray}
where $m\in \left[0,\infty\right)$.\\ Some previous studies have demonstrated that equations of the type given in Eq. (\ref{eq:9}) can be used to model light dynamics in the so-called Glauber-Fock photonic lattices \cite{GF3}. Unlike normal Glauber-Fock lattices, in the present case we have an extra term $(-1)^{m}$, which indicates that these lattices must be endowed with alternating positive and negative coupling coefficients apart from the square root law distribution \cite{GF,blas}.\\
Note that in the quantum realm $E_m(z)=\langle{m}|\exp\left(iz\hat{H}\right)|{\Psi\left(0\right)}\rangle$ represents the transition probability amplitude for a system initially prepared in state $|{\Psi\left(0\right)}\rangle$ to populate the number state  $|{m}\rangle$. In the classical case of Glauber-Fock arrays, $E_m(z)$ represents the optical field amplitude at waveguide $m$ after a propagation distance $z$ when the initial field distribution $|{\Psi\left(0\right)}\rangle$ is launched at $z=0$ \cite{GF, GF3}. To experimentally realize such photonic arrays, it is necessary to impose a square root law distribution plus an extra phase shift in the coupling coefficients. The physical realization of negative couplings is nowadays possible by inducing a longitudinal modulation of the refractive index of the waveguides as indicated in reference \cite{Zeuner2}.\\
Direct integration of Eq. (\ref{eq:8}) renders the lattice evolution operator
\begin{eqnarray}\label{eq:10}
U(z)=\exp\left(-igz[\hat{a}^{\dagger}(-1)^{\hat{n}}+(-1)^{\hat{n}}\hat{a}]\right),
\end{eqnarray}
which is identical to the deformed displacement operator given in Eq. (\ref{eq:1}) provided $\alpha=-igz$. In order to explore the light dynamics in these types of arrays, we consider the Green function $E_{m}\left(z\right)=\langle{m}|U(z)|{k}\rangle$, which gives the field amplitude at channel $m$ upon excitation of site $k$. For sites laying to the left of the excited channel, $m\leq k$, we obtain the expressions
\begin{eqnarray}
\label{eq:11}
&& E_{m}\left(z\right)=
\frac{1}{\sqrt{2}}\sqrt{\frac{m!}{k!}}\exp\left(-\theta^2/2\right)L_{m}^{k-m}\left(\theta^{2}\right)\nonumber \\ 
&&\times \left\{ \begin{array}{ll}
\exp(-i\pi/4) \theta^{k-m}+ \exp(i\pi/4) \left(-\theta\right)^{k-m},\quad k = odd,\\\exp(i\pi/4) \theta^{k-m}+ \exp(-i\pi/4) \left(-\theta\right)^{k-m},\quad k = even.\end{array}\right.,
\end{eqnarray}
whereas for for sites at the right, $m\geq k$, we have
\begin{eqnarray}
\label{eq:12}
&& E_{m}\left(z\right)=
\frac{1}{\sqrt{2}}\sqrt{\frac{k!}{m!}}\exp\left(-\theta^2/2\right)L_{k}^{m-k}\left(\theta^{2}\right)\nonumber \\  &&\times \left\{ \begin{array}{ll}
\exp(i\pi/4)\theta^{m-k}+\exp(-i\pi/4)\left(-\theta\right)^{m-k}, \quad k = odd,\\\exp(-i\pi/4)\theta^{m-k}+\exp(i\pi/4)\left(-\theta\right)^{m-k},\quad k = even.\end{array}\right.
\end{eqnarray}
Here $\theta=gz$ and $L_{k}^{m}(\theta)$ represents the associated Laguerre polynomials. A detailed derivation of Eq. (\ref{eq:11})  and Eq. (\ref{eq:12}) is given in the Appendix.\\ 
In a normal non-deformed Glauber-Fock photonic lattice \cite{GF,GF2,GF3}, excitation of the $0$-th waveguide   produces intensity patterns reminiscent to the probability distribution of coherent states. Moreover, when exciting any other site $|{k}\neq |{0}$, the intensity resembles the probability distribution of the $k$-th displaced Fock state \cite{Oliveira}. Correspondingly, in the present case, Eq. (\ref{eq:6a}) (Eq.\ref{eq:7}) indicates that excitation of the $"vacuum"$ state $|{0}\rangle$ ($"number"$ state $|{k}\rangle$) will produce field amplitudes which are build up of a superposition of two Poisson (displaced Fock) distributions in the waveguide number. The Poisson superpositions become apparent by considering $k=0$ in the analytic solutions Eq. (\ref{eq:11}) and Eq. (\ref{eq:12})
\begin{eqnarray}\label{eq:13}
E_{m}(z)=\frac{\exp(-i\pi/4)}{\sqrt{2}}\exp\left(-\frac{\left(gz\right)^{2}}{2}\right)\frac{\left(gz\right)^{m}}{\sqrt{m!}}\nonumber \\+\frac{\exp(i\pi/4)}{\sqrt{2}}\exp\left(-\frac{\left(gz\right)^{2}}{2}\right)\frac{\left(-gz\right)^{m}}{\sqrt{m!}}.
\end{eqnarray}
To illustrate these effects, in Figure (\ref{Fig1}.a) we show the theoretically calculated intensity evolution when the $0th$ waveguide is initially excited in a deformed Glauber-Fock lattice of 60 waveguides and $g=1$. For the case when any other site $k\neq 0$ is excited, the field amplitudes become a superposition of two "displaced Fock states" \cite{Oliveira} separated in phase, the field amplitudes are given by Eq. (\ref{eq:11}) and Eq. (\ref{eq:12}). In  Fig.(\ref{Fig1}.c) we present the propagation dynamics when the third "Fock state" is initially excited.\\ 
In the previous section we pointed out that application of the deformed displacement operator to DFSs produces high order superpostions of DFSs. As a result, deformed Glauber-Fock lattices are expected to give rise optical field profiles entirely analogous to such DFSs superpositions, provided we excitate the system with discrete wave packets featuring amplitudes akin to DFSs, $|{\psi(0)}\rangle=|{\beta,k}\rangle$. 
Hence, assuming such initial field for deformed Glauber-Fock lattices we obtain a superposition of four DFSs with equal amplitude and different phase
\begin{eqnarray}\label{eq:14}
|{\psi\left(z\right)}\rangle=\frac{\exp\left(igz Im\left(\beta\right)\right)}{2}\left(|{-gZ+\beta,k}\rangle-i\left(-1\right)^{k}|{gz-\beta,k}\rangle\right)\nonumber\\+\frac{\exp\left(-igz Im\left(\beta\right)\right)}{2}\left(|{gz+\beta,k}\rangle+i\left(-1\right)^{k}|{-gz-\beta,k}\rangle\right),
\end{eqnarray}
where $Im\left(\beta\right)$ indicates imaginary part, see Eq. (\ref{eq:7}). 
Similarly, for the case when the lattice is excited with a "coherent state" wavepacket, $|{\psi(0)}\rangle=|{\beta}\rangle$, Eq. (\ref{eq:14}) reduces to a superposition of four "coherent states"
\begin{eqnarray}\label{eq:15}
|{\psi\left(z\right)}\rangle=\frac{\exp\left(igz Im\left(\beta\right)\right)}{2}\left(|{-gz+\beta}\rangle-i|{gz-\beta}\rangle\right)\nonumber\\+\frac{\exp\left(-igz Im\left(\beta\right)\right)}{2}\left(|{gz+\beta}\rangle+i|{-gz-\beta}\rangle\right).
\end{eqnarray}
In Figs.(\ref{fig:Fig2}) and (\ref{fig:Fig3}) we show the intensity evolution for a coherent and a displaced Fock state, respectively. In both cases, we observe two counterpropagating beams each of which are build up of two CS (Fig.~(\ref{fig:Fig2})) and two DFS (Fig.~(\ref{fig:Fig3})), as predicted by Eq. (\ref{eq:14}) and Eq. (\ref{eq:15}). Certainly, the splitting of the initial CS (DFS) into four CS (DFS) can be thought of as optical analogues of generalized Schr\"odinger cat states.

\section{Wigner functions in deformed Glauber-Fock lattices}

\begin{figure}[t]
\centerline{\includegraphics[scale=0.3]{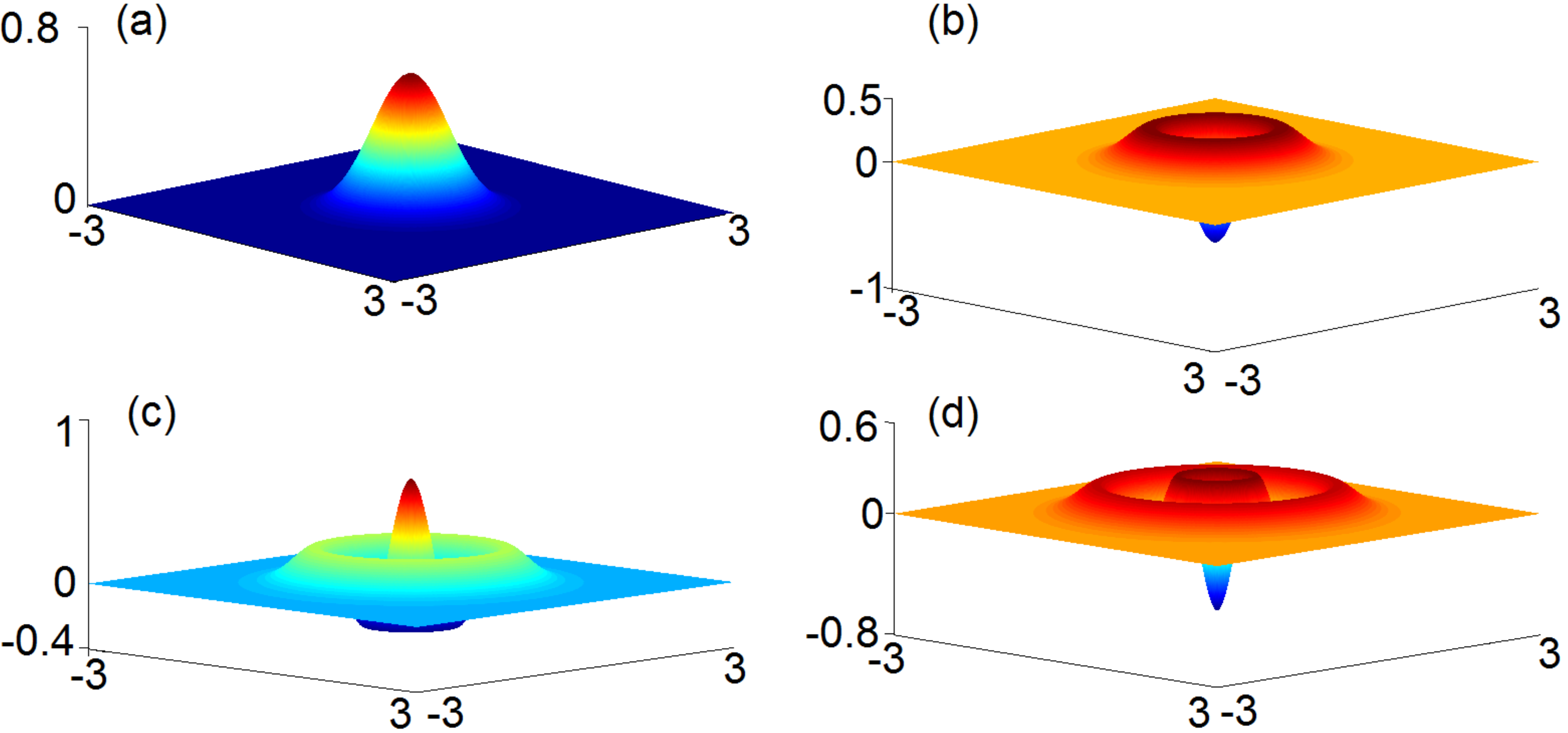}}
\caption{Wigner functions for (a) the vacuum state $|{0}\rangle$, (b) the first $|{1}\rangle$, (c) the second $|{2}\rangle$, and (d) the third $|{3}\rangle$ Fock states. For a coherent state the Wigner function is the same as for the vacuum state (a), but displaced to any of the quadrants of the phase space.}\label{fig:Fig4}
\end{figure}

\begin{figure}[h!]
\includegraphics[scale=0.3]{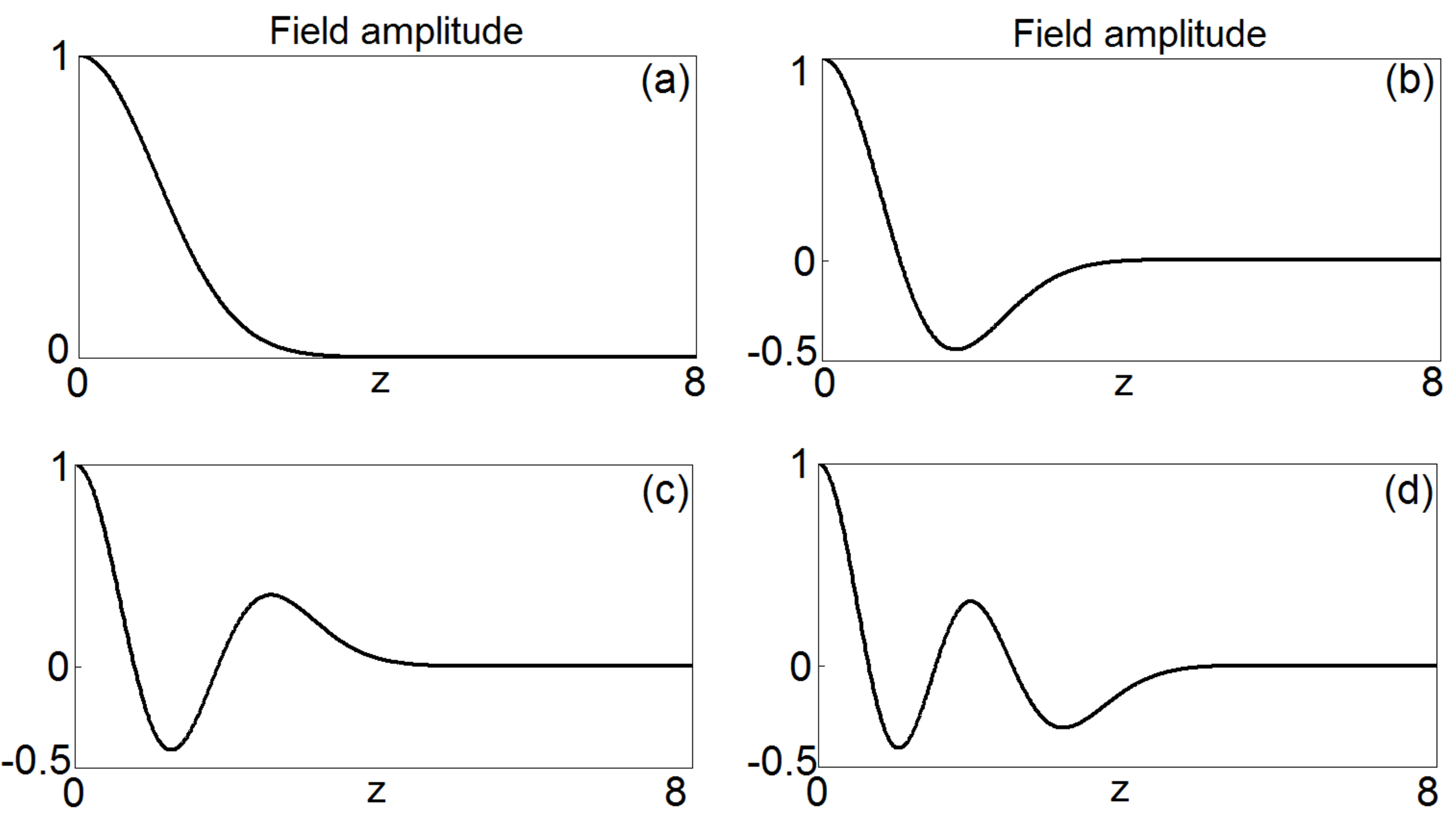}
\caption {Light evolution along the first four waveguides of a deformed Glauber-Fock lattice with 160 waveguides and $g=1$. In (a) the field amplitude along the $0$-th waveguide reproduces part of the Wigner function corresponding to the vacuum state. Similarly, in (b), (c), and (d) we depict sections of the Wigner functions corresponding to the first, second, and third Fock state, respectively. In all the cases the initial field amplutude is unitary and the propagation distance is given in centimeters.}\label{fig:Fig5}
\end{figure}

Among many representations of quantum states, Wigner functions offer the possibility of describing nonclassicality using the well-established concept of phase space \cite{Gerry}. In this regard, here we show that Wigner functions of Fock states and CS can be observed by using the evolution operator associated to the Hamiltonian $\hat{H}=g\left(\hat{a}^{\dagger}(-1)^{\hat{n}} + (-1)^{\hat{n}} \hat{a}\right)$. As a result, these effects can be emulated classically in deformed Glauber-Fock lattices. \\
We begin by introducing the Wigner function for an arbitrary pure state $|{\psi}$ at a complex phase-space point $\alpha=x+iy$ \cite{Wigner1,Hillery,Moya}
\begin{equation}\label{eq:16}
W(\alpha)=\langle {\psi}|\hat{D}(2\alpha)(-1)^{\hat{n}}|{\psi}\rangle,
\end{equation}
where we have dropped the trivial factor $\pi/2$. It is worth noting that this representation of the Wigner function has been thoroughly used for the reconstruction of vibrational motion of ion systems \cite{Wineland2} as well as to retrieve quantized optical fields \cite{Bertet}. Using Eq. (\ref{eq:16}) one can  show that the Wigner function of pure Fock states, $|{\psi}\rangle=|{k}\rangle$, is given by \cite{Gerry, Moya}
\begin{equation}\label{eq:17}
W(\alpha)=(-1)^{k}\exp\left(-2|\alpha|^{2}\right)L_{k}\left(4|\alpha|^{2}\right).
\end{equation}
with $L_{k}\left(4|\alpha|^{2}\right)$ representing the Laguerre polynomials of order $k$. Correspondingly for a coherent state $|{\beta}\rangle$ we obtain
\begin{equation}\label{eq:18}
W(\alpha)=\exp\left(-2|\beta-\alpha|^{2}\right).
\end{equation}
For illustrative purposes in Fig.(\ref{fig:Fig4}) we present Wigner functions  for the vacuum and the first three Fock states. 
We now turn our attention to the evolution operator of our deformed Glauber-Fock lattices Eq. (\ref{eq:10}) acting over the initial state $|{k}\rangle$, and monitoring the light dynamics along the same waveguide
\begin{eqnarray}\label{eq:19}
\langle{k}|U(z)|{k}\rangle=\frac{1}{2}\left[\langle{k}|\hat{D}\left(gz\right)|{k}\rangle+\langle{k}|\hat{D}^{\dagger}\left(gz\right)|{k}\rangle\right]\nonumber \\ -\frac{i}{2}\left(-1\right)^{k}\left[\langle{k}|\hat{D}\left(gz\right)|{k}\rangle-\langle{k}|\hat{D}^{\dagger}\left(gz\right)|{k}\rangle\right].
\end{eqnarray} 
A direct calculation reveals that the expectation value $\langle{k}|\hat{D}\left(gz\right)|{k}\rangle=\langle{k}|\hat{D}^{\dagger}\left(gz\right)|{k}\rangle$, and as a result, Eq. (\ref{eq:19}) becomes
\begin{eqnarray}\label{eq:20}
\langle{k}|U(z)|{k}\rangle=\exp\left(-\frac{\left(gz\right)^{2}}{2}\right)L_{k}\left(\left(gz\right)^{2}\right).
\end{eqnarray}
This indicates that injecting light into channel $|{k}\rangle$ of a deformed Glauber-Fock lattice and monitoring the field evolution along the same waveguide we emulate the Wigner function of the "number state" $|{k}\rangle$ including the phase. Comparing Eq. (\ref{eq:17}) and Eq. (\ref{eq:20}) we see that the only difference is the factor $(-1)^k$. As a consequence, Eq. (\ref{eq:20}) describes Wigner functions of even number states, while for odd number states it has to be multiplied by -1.
Fig. (\ref{fig:Fig5}) depicts the light evolution along the $0$-th, first, second, and third waveguides when these same wavegides are initially excited. These plots correspond to one line of the Wigner functions for the vacuum and the first three Fock states. 
\section{Conclusion}
We demonstrate that superpositions of coherent and displaced Fock states, also referred to as generalized Schr\"odinger cats, can be created by application of a nonlinear displacement operator- a deformed version of the Glauber displacement operator. Consequently, such generalized cat states can be formally considered as nonlinear coherent states. We then show that Glauber-Fock photonic lattices endowed with alternating positive and negative coupling coefficients give rise to classical analogs of such cat states. In addition, it is pointed out that the analytic propagator of these deformed Glauber-Fock arrays explicitly contains the Wigner operator opening the possibility to observe Wigner functions of the quantum harmonic oscillator in the classical domain.
\section{Appendix}
In order to compute the Green function of the Glauber Fock lattice we start from the evolution operator Eq. (\ref{eq:10})
\begin{eqnarray}\label{eq:21}
U(z)=\exp(-igz[\hat{a}^{\dagger}(-1)^{\hat{n}}+(-1)^{\hat{n}}\hat{a}]).
\end{eqnarray}
This operator can be easily disentangled as follows
\begin{eqnarray}\label{eq:22}
U(z)=\exp\left(-\frac{(gz)^2}{2}\right) \exp\left(-igz \hat{a}^{\dagger}(-1)^{\hat{n}}\right) \exp\left(-igz(-1)^{\hat{n}}\hat{a}\right).
\end{eqnarray}
Developing the exponentials in even and odd terms we obtain
\begin{eqnarray}\nonumber
U(z)&=&\exp\left(-\frac{(gz)^2}{2}\right)  \left[\sum_k\frac{(-igz)^{2k}}{(2k)!}(\hat{a}^{\dagger}(-1)^{\hat{n}})^{2k}+\sum_k\frac{(-igz)^{2k+1}}{(2k+1)!}(\hat{a}^{\dagger}(-1)^{\hat{n}})^{2k+1} \right]\nonumber \\
&\times&\left[\sum_k\frac{(-igZ)^{2k}}{(2k)!}((-1)^{\hat{n}}\hat{a})^{2k}+\sum_k\frac{(-igz)^{2k+1}}{(2k+1)!}((-1)^{\hat{n}}\hat{a})^{2k+1} \right].\nonumber
\end{eqnarray}
And using the identities
\begin{eqnarray}\nonumber
((-1)^{\hat{n}}\hat{a})^{2k}=(-1)^{k}\hat{a}^{2k},\nonumber\\
((-1)^{\hat{n}}\hat{a})^{2k+1}=(-1)^{k}(-1)^{\hat{n}}\hat{a}^{2k+1},\nonumber\\
(\hat{a}^{\dagger}(-1)^{\hat{n}})^{2k}=(-1)^{k}\hat{a}^{\dagger 2k},\nonumber\\
(\hat{a}^{\dagger}(-1)^{\hat{n}})^{2k+1}=(-1)^{k}\hat{a}^{\dagger 2k+1}(-1)^{\hat{n}}\nonumber,
\end{eqnarray}
we can write the evolution operator in terms of hyperbolic functions
\begin{eqnarray}\nonumber
U(z)&=&\exp\left(-\frac{\left(gz\right)^{2}}{2}\right) [\cosh\left(gz\hat{a}^{\dagger}\right)-i\sinh\left(gz\hat{a}^{\dagger}\right)(-1)^{\hat{n}}]\\&\times&[\cosh\left(gz\hat{a}\right)-i(-1)^{\hat{n}}\sinh(gz\hat{a})].\nonumber
\end{eqnarray}
Developing the hyperbolic function in terms of exponentials and noting that $(1-i(-1)^{\hat{n}})(1+i(-1)^{\hat{n}})=2$ and $(1-i(-1)^{\hat{n}})^2=-2i(-1)^{\hat{n}}$, the evolution operator becomes
\begin{eqnarray}\nonumber
U(z)=\exp\left(-\frac{\left(gz\right)^{2}}{2}\right)\left[\exp\left(gz\hat{a}^{\dagger}\right) \frac{1-i(-1)^{\hat{n}}}{2} +\exp\left(-gz\hat{a}^{\dagger}\right)\frac{1+i(-1)^{\hat{n}}}{2}\right]\\
\left[\frac{1-i(-1)^{\hat{n}}}{2}\exp\left(gz\hat{a}\right)+\frac{1+i(-1)^{\hat{n}}}{2}\exp\left(-gz\hat{a}\right)\right].\nonumber
\end{eqnarray}
Rearranging terms  and using the fact that $(-1)^{\hat{n}}f(\hat{a})=f(-\hat{a})(-1)^{\hat{n}}$ we obtain
\begin{eqnarray}\nonumber
U(z)&=&\exp\left(-\frac{(gz)^2}{2}\right)\left[\exp\left(gz\hat{a}^{\dagger}\right) \exp\left(-gz\hat{a}\right)\frac{1-i(-1)^{\hat{n}}}{2}\right.\\
&+&\left.\exp\left(-gz\hat{a}^{\dagger}\right)\exp\left(gz\hat{a}\right)\frac{1+i(-1)^{\hat{n}}}{2}\right].\label{eq:23}
\end{eqnarray}
In the optical context of Glauber-Fock lattices the Green function describes the field amplitude at the m-th waveguide when the k-th site is excited and is given mathematically by the expectation value $E_{m}(z)=\langle{m}|U(z)|{k}\rangle$, with $U(z)$ given in Eq. (\ref{eq:23}). Computation of the Green function is easily done using the expansions 
\begin{eqnarray}\nonumber
\exp\left(-gz\hat{a}\right)(-1)^{\hat{n}}|{k}=(-1)^{k}\sum_{l=0}^{k}\frac{(-gz)^{l}}{l!}\sqrt{\frac{k!}{(k-l)!}}|{k-l}, \qquad for \qquad l\leq k
\end{eqnarray}
and
\begin{eqnarray}\nonumber
\langle{m}|\exp\left(gz\hat{a}^{\dagger}\right)=\sum_{n=0}^{m}\frac{(gz)^{n}}{n!}\sqrt{\frac{m!}{(m-n)!}}\langle{m-n}|, \qquad for \qquad n\leq m
\end{eqnarray}
which altogether give
\begin{equation}\label{eq:24}
\langle{m}|\exp\left(gz\hat{a}^{\dagger}\right)\exp\left(-gz\hat{a}\right)(-1)^{\hat{n}}|{k}\rangle=(-1)^{k}\sum_{n,l=0}^{m,k}\frac{(gz)^{n+l}(-1)^{l}}{n!l!}\sqrt{\frac{m!k!}{(m-n)!(k-l)!}}\delta_{m-n,k-l}.
\end{equation}
By considering $n=m+l-k$ we obtain
\begin{equation}\label{eq:25}
\langle{m}|\exp\left(gz\hat{a}^{\dagger}\right)\exp\left(-gz\hat{a}\right)(-1)^{\hat{n}}|{k}\rangle=(-1)^{k}\sum_{l=0}^{k}\frac{(gz)^{2l+m-k}(-1)^{l}\sqrt{m!k!}}{l!(m+l-k)!(k-l)!}.
\end{equation}
Now, assuming $m=k+s$ yields
\begin{equation}\label{eq:26}
\langle{k+s}|\exp\left(gz\hat{a}^{\dagger}\right)\exp\left(-gz\hat{a}\right)(-1)^{\hat{n}}|{k}\rangle=(-1)^{k}\left(gz\right)^{s}\sqrt{\frac{k!}{(k+s)!}}\sum_{l=0}^{k}\frac{\left((gz)^2\right)^{l}(-1)^{l}(k+s)!}{l!(s+l)!(k-l)!}. 
\end{equation}
From this expression we recognize the associated Laguerre polynomials
\begin{equation}
L_{k}^{s}\left((gz)^2\right)=\sum_{l=0}^{k}\frac{\left((gz)^2\right)^{l}(-1)^{l}(k+s)!}{l!(s+l)!(k-l)!}.
\label{eq:27}
\end{equation}
Thus, Eq. (\ref{eq:26}) becomes
\begin{eqnarray}\label{eq:28}
\langle{k+s}|\exp\left(gz\hat{a}^{\dagger}\right)\exp\left(-gz\hat{a}\right)(-1)^{\hat{n}}|{k}\rangle=(-1)^{k}\left(gz\right)^{s}\sqrt{\frac{k!}{(k+s)!}}L_{k}^{s}\left((gz)^2\right)
\end{eqnarray}
which for $s=m-k$ 
\begin{eqnarray}\label{eq:29}
\langle{m}|\exp\left(gz\hat{a}^{\dagger}\right)\exp\left(-gz\hat{a}\right)(-1)^{\hat{n}}|{k}\rangle=(-1)^{k}\left(gz\right)^{m-k}\sqrt{\frac{k!}{m!}}L_{k}^{m-k}\left((gz)^2\right)
\end{eqnarray}
This last expression is valid for all the waveguides $m\geq k$. Returning to Eq. (\ref{eq:24}), considering $l=n+k-m$, then $m=k-s$, and assuming $s=k-m$ we obtain 
\begin{eqnarray}\label{eq:30}
\langle{m}|\exp\left(gz\hat{a}^{\dagger}\right)\exp\left(-gz\hat{a}\right)(-1)^{\hat{n}}|{k}\rangle=(-1)^{k}\left(-gz\right)^{k-m}\sqrt{\frac{m!}{k!}}L_{m}^{k-m}\left((gz)^2\right)
\end{eqnarray}
which is valid for all the waveguides $m\leq k$. In the same fashion we can obtain the other terms in the Green function, which for the sake of clarity are summarized here
\begin{eqnarray}\label{eq:31}
\langle{m}|\exp\left(-gz\hat{a}^{\dagger}\right)\exp\left(gz\hat{a}\right)(-1)^{\hat{n}}|{k}\rangle=(-1)^{k}\left(-gZ\right)^{m-k}\sqrt{\frac{k!}{m!}}L_{k}^{m-k}\left((gz)^2\right),\quad m\geq k
\end{eqnarray}
\begin{equation}\label{eq:32}
\langle{m}|\exp\left(-gz\hat{a}^{\dagger}\right)\exp\left(gz\hat{a}\right)(-1)^{\hat{n}}|{k}\rangle=(-1)^{k}\left(gz\right)^{k-m}\sqrt{\frac{m!}{k!}}L_{m}^{k-m}\left((gz)^2\right),\quad m\leq k
\end{equation}
\begin{equation}\label{eq:33}
\langle{m}|\exp\left(-gz\hat{a}^{\dagger}\right)\exp\left(gz\hat{a}\right)|{k}\rangle=\left(-gz\right)^{m-k}\sqrt{\frac{k!}{m!}}L_{k}^{m-k}\left((gz)^2\right),\quad m\geq k
\end{equation}
\begin{equation}\label{eq:34}
\langle{m}|\exp\left(-gz\hat{a}^{\dagger}\right)\exp\left(gz\hat{a}\right)|{k}\rangle=\left(gz\right)^{k-m}\sqrt{\frac{m!}{k!}}L_{m}^{k-m}\left((gz)^2\right),\quad m\leq k
\end{equation}
\begin{equation}\label{eq:35}
\langle{m}|\exp\left(gz\hat{a}^{\dagger}\right)\exp\left(-gz\hat{a}\right)|{k}\rangle=\left(gz\right)^{m-k}\sqrt{\frac{k!}{m!}}L_{k}^{m-k}\left((gz)^2\right),\quad m\geq k
\end{equation}
\begin{equation}\label{eq:36}
\langle{m}|\exp\left(gz\hat{a}^{\dagger}\right)\exp\left(-gz\hat{a}\right)|{k}\rangle=\left(-gz\right)^{k-m}\sqrt{\frac{m!}{k!}}L_{m}^{k-m}\left((gz)^2\right),\quad m\leq k
\end{equation}
Finally, using these expressions we obtain the Green function given in Eq. (\ref{eq:11}) and Eq. (\ref{eq:12}).


\begin{thebibliography}{XX}

\bibitem{Haroche}  S. Haroche, Rev. Mod. Phys. 85, 1083 (2013).
\bibitem{Wineland1} D. J. Wineland, Rev. Mod. Phys. 85, 1103 (2013).
\bibitem{Silberberg} I. Afek, O. Ambar, and Y. Silberberg, Science 328, 879 (2010).
\bibitem{Gerry} C. C. Gerry and P. L. Knight, Introductory quantum optics (Cambridge University Press,
2005).
\bibitem{Heilmann} R. Heilmann, J. Sperling, A. Perez-Leija, M. Graefe, M. Heinrich, S. Nolte, W. Vogel,
and A. Szameit, ArXiv:1502.04932.
\bibitem{Manko} V. I. Man’ko, G. Marmo, E. C. G. Sudarshan, and F. Zaccaria, Physica Scripta 55, 528
(1997).
\bibitem{Glauber1} R. J. Glauber, Phys. Rev. 130, 2529 (1963).
\bibitem{Glauber2} R. J. Glauber, Phys. Rev. Lett. 10, 84 (1963).
\bibitem{Schrodinger} E. Schr\"odinger, Naturwissenschaften 23, 823 (1935).
\bibitem{Yurke} B. Yurke and D. Stoler, Phys. Rev. Lett. 57, 13 (1986).
\bibitem{GF} A. Perez-Leija, H. Moya-Cessa, A. Szameit, and D. N. Christodoulides, Opt. Lett. 35,
2409 (2010).
\bibitem{louisell} W. H. Louisell, Quantum and statistical properties of radiation (Wiley, 1973).
\bibitem{Gerry2} C. C. Gerry and J. Mimih, Contemporary Phys. 51, 497 (2010).
\bibitem{Christodoulides} D. N. Christodoulides, F. Lederer, and Y. Silberberg, Nature 424, 817 (2003).
\bibitem{Review} T. Meany, M. Gr\"afe, R. Heilmann, A. Perez-Leija, S. Gross, M. J. Steel, M. J. Withford,
and A. Szameit, Laser and Photonics Reviews 9, 363 (2015).
\bibitem{Garanovich} I. L. Garanovich, A. Szameit, A. A. Sukhorukov, T. Pertsch, W. Krolikowski, S. Nolte,
D. Neshev, A. Tuennermann, and Y. S. Kivshar, Opt. Express 15, 9737 (2007).
\bibitem{Wstates} A. Perez-Leija, J. C. Hernandez-Herrejon, H. Moya-Cessa, A. Szameit, and D. N.
Christodoulides, Phys. Rev. A 87, 013842 (2013).
\bibitem{Longhi} S. Longhi, Opt. Lett. 38, 4884 (2013).
\bibitem{GF2} A. Perez-Leija, R. Keil, A. Szameit, A. F. Abouraddy, H. Moya-Cessa, and D. N.
Christodoulides, Phys. Rev. A 85, 013848 (2012). 16
￼
\bibitem{GF3} R. Keil, A. Perez-Leija, F. Dreisow, M. Heinrich, H. Moya-Cessa, S. Nolte, D. N. Christodoulides, and A. Szameit, Phys. Rev. Lett. 107, 103601 (2011).
\bibitem{PT} K. G. Makris, R. El-Ganainy, D. N. Christodoulides, and Z. H. Musslimani, Phys. Rev. Lett. 100, 103904 (2008).
\bibitem{Kai} K. Wang, S. Weimann, S. Nolte, A. Perez-Leija, and A. Szameit, ArXiv:1510.08601v1.
\bibitem{Zeuner} J. M. Zeuner, N. K. Efremidis, R. Keil, F. Dreisow, D. N. Christodoulides,
A. Tu ̈nnermann, S. Nolte, and A. Szameit, Phys. Rev. Lett. 109, 023602 (2012).
\bibitem{Efremidis} N. K. Efremidis, P. Zhang, Z. Chen, D. N. Christodoulides, C. E. Ru ̈ter, and D. Kip,
Phys. Rev. A 81, 053817 (2010).
\bibitem{Metcalf} B. J. Metcalf, J. B. Spring, P. C. Humphreys, N. Thomas-Peter, M. Barbieri, W. S.
Kolthammer, X.-M. Jin, N. K. Langford, D. Kundys, J. C. Gates, B. J. Smith, P. G. R.
Smith, and I. A. Walmsley, Nature Photonics 8, 770 (2014).
\bibitem{Wstates2} M. Gr\"afe, R. Heilmann, A. Perez-Leija, R. Keil, F. Dreisow, M. Heinrich, H. Moya-Cessa,
S. Nolte, D. N. Christodoulides, and A. Szameit, Nature Photonics 8, 791 (2014).
\bibitem{Jx1} A. Perez-Leija, R. Keil, A. Kay, H. Moya-Cessa, S. Nolte, L.-C. Kwek, B. M. Rodriguez-
Lara, A. Szameit, and D. N. Christodoulides, Phys. Rev. A 87, 012309 (2013).
\bibitem{Jx2} A. Perez-Leija, R. Keil, H. Moya-Cessa, A. Szameit, and D. N. Christodoulides, Phys.
Rev. A 87, 022303 (2013).
\bibitem{Fourier} S. Weimann, A. Perez-Leija, M. Lebugle, R. Keil, M. Tichy, M. Grfe, R. Heilmann, S. Nolte, H. Moya-Cessa, G. Weihs, D. N. Christodoulides, and A. Szameit,
ArXiv:1508.00033.
\bibitem{Peruzzo} A. Peruzzo, M. Lobino, J. C. F. Matthews, N. Matsuda, A. Politi, K. Poulios, X.-Q.
Zhou, Y. Lahini, N. Ismail, K. Wrhoff, Y. Bromberg, Y. Silberberg, M. G. Thompson,
and J. L. O’Brien, Science 329, 1500 (2010).
\bibitem{blas} B. M. Rodriguez-Lara, Phys. Rev. A 84, 053845 (2011).
\bibitem{Zeuner2} J. M. Zeuner, M. C. Rechtsman, R. Keil, F. Dreisow, A. T\"unnermann, S. Nolte, and
A. Szameit, Opt. Lett. 37, 533 (2012).
\bibitem{Oliveira} F. A. M. de Oliveira, M. S. Kim, P. L. Knight, and V. Buek, Phys. Rev. A 41, 2645
(1990).
\bibitem{Wigner1} E. Wigner, Phys. Rev. 40, 749 (1932).
\bibitem{Hillery} M. Hillery, R. O’Connell, M. Scully, and E. Wigner, Physics Reports 106, 121 (1984).
17
\bibitem{Moya} H. Moya-Cessa and P. L. Knight, Phys. Rev. A 48, 2479 (1993).
\bibitem{Wineland2} D. Leibfried, D. M. Meekhof, B. E. King, C. Monroe, W. M. Itano, and D. J. Wineland,
Phys. Rev. Lett. 77, 4281 (1996).
\bibitem{Bertet} P. Bertet, A. Auffeves, P. Maioli, S. Osnaghi, T. Meunier, M. Brune, J. M. Raimond,
and S. Haroche, Phys. Rev. Lett. 89, 200402 (2002).
\end{thebibliography}
\end{document}